\documentclass[a4paper,12pt]{article}

\usepackage{amsmath}
\usepackage{hyperref}
\usepackage{cite}
\usepackage{amssymb}
\usepackage{braket}

\numberwithin{equation}{section}

\title{Symmetry Structure of ADHM Sigma Models and $AdS_3$ Superstrings}
\author{Abbas Ali\footnote{Email:aali.ph@gmail.com}, Mohsin Ilahi, P.P. Abdul Salih \\and Shafeeq Rahman Thottoli\footnote{Present Address : Department of Physical Sciences, Physics Division, College of Science,
Jazan University, Jazan 45142, Kingdom of Saudi Arabia}\\
		Physics Department,\\
  Aligarh Muslim University,\\ Aligarh-202 002, India}
\date{}
\begin{document}

\maketitle

\begin{abstract}

We clarify the supersymmetry structure of  ADHM instanton linear sigma models, $N=4$ superconformal algebras and superconformal field theories dual to superstrings moving in $AdS_3$ backgrounds in view of some of our recent findings in various investigations. First of all we collect information about the (0, 4) supersymmetries of Witten's original, Ali-Ilahi's complementary   and Ali-Salih's complete ADHM instanton linear sigma models. Then we collect information about large, small and middle $N=4$ superconformal algebras and summarize  their interrelations  with the ADHM instanton linear sigma models.  We then argue that the former, that is the supersymmetry of the ADHM instanton linear sigma models, flows to latter, that is the small, middle and large $N=4$ superconformal symmetries, in the infrared limit. Finally we summarize the mapping of these $N=4$ superconformal symmetries onto the superconformal symmetries dual to superstrings moving on $AdS_3\times S^3\times {\mathcal M}^4$ backgrounds with ${\mathcal M}^=K3, T^4$ and $S^3\times S^1$.

\end{abstract}

\newpage

\tableofcontents

\section{Introduction}

$AdS_5$ and $AdS_3$ are the two most intensely studied cases of the famous Maldacena Conjecture about duality between theories with gravity on one side and conformal field theories without gravity on the other \cite{Maldacena:1997re, Witten:1998qj, Gubser:1998bc}. In the case of $AdS_3$ three distinct  cases have been investigated very extensively. These are the cases of Type IIB superstrings moving on manifolds with geometry $AdS_3 \times S^3 \times \mathcal{M}^4$ where $\mathcal{M}^4= T^4, K3~~\text{or}~~S^3 \times S^1$. This case is a prime target in the saga of AdS/CFT investigation because of the fact that in this case the dual conformal field theory happens to be two dimensional and hence the whole formidable machinery of two dimensional conformal field theory becomes available to us. 

Evidence so far uncovered points to the fact that the field theory dual to superstrings  moving on $AdS_3 \times S^3  \times K3$ background geometry have the small N=4  super conformal symmetry\cite{Ademollo:1975an, Giveon:1998ns, David:2002wn} while for superstrings moving on $AdS_3 \times S^3  \times T^4$ geometry it is the middle N=4 superconformal symmetry \cite{Ali:1993sd, Hasiewicz:1989vp, Ali:2003aa, Ali:2000we}. This has been put to good use and we have proofs and derivations of the correspondence in this case\cite{Giveon:1998ns, Eberhardt:2019ywk}. The field theory dual to superstrings moving on $AdS_3 \times S^3\times S^3 \times S^1$ geometry has  the large $N=4$\cite{SEVRIN1988447, Ivanov:1988rt, Schoutens:1988ig, Ali:2000zu} superconformal symmetry.

Though this all is a legend by now but proofs in some cases are anecdotal and lack derivations. In the present note we present a narrative that should convert this legend into logic. We shall take advantage of our work on ADHM instanton linear sigma models to make the symmetry structure of $AdS_3$ superstrings more clear.
 
Maldacena conjectured that the two dimensional superconformal field theory describing the Higgs branch of D1-D5 system on $M^4$ is dual to type IIB superstring theory moving on $AdS_3 \times S^3  \times M^4$ with $M^4= K3$ or $T^4$. This  Higgs branch has small $N=4$  superconformal symmetry in case of $AdS_3 \times S^3  \times K3$ and middle $N=4$ superconformal symmetry in $AdS_3 \times S^3  \times M^4$ case. This Higgs branch has been analysed in Refs.\cite{Strominger:1996sh, Vafa:1995zh, Vafa:1995bm, Dijkgraaf:1996cv, Dijkgraaf:1996hk, Dijkgraaf:1997nb, Hassan:1997ai, Maldacena:1998bw}. Corresponding superconformal field theory is the symmetric orbifold theory of the compact part of the manifold. Related Coulonmb branch was analysed in Ref.\cite{Diaconescu:1997gu}.

Similar identification of the dual superconformal field theory should go through for $AdS_3 \times S^3\times S^3 \times S^1$ superstrings - that the corresponding dual superconformal field theory is the symmetric orbifold theory of the compact manifold $S^3 \times S^1$. Different efforts, Ref.\cite{Elitzur:1998mm, deBoer:1999gea, Gukov:2004ym, Tong:2014yna, Eberhardt:2017fsi, Eberhardt:2017pty, Papadopoulos:2024uvi, Witten:2024yod}, bring us tantalizingly close to this conclusion but not completely so. In view of the fact that corresponding investigations are very thorough it means that we are missing something very basic in this case. 

The problem is all the more surprising because it happens to be related a number of very well investigated stringy systems like two dimensional sigma models \cite{Howe:1987qv, Howe:1988cj}, 't Hooft instanton sigma models and NS1-NS5 brane system  \cite{Strominger:1990et, Callan:1991dj, Callan:1991ky, Callan:1991at, Dabholkar:1990yf, Tseytlin:1996as}, ADHM instanton sigma models\cite{Witten:1994tz, Ali:2023csc, Ali:2023ams, Ali:2023icn, Lambert:1995dp},  D1-D5 brane system\cite{Callan:1996dv, Horowitz:1996ay}, matrix models\cite{Dijkgraaf:1997nb, Aharony:1997th}, BTZ black hole and its stringy generalization\cite{Banados:1992wn, Ali:1992mj}. In our investigations we are examining this problem from some of these angles. In Ref.\cite{Ali:2023kkf} we took this problem from the point of view of free field realizations of the large $N=4$ superconformal algebra and presented a complete free field realization of the large $N=4$ superconformal algebra. We believe that this free field realization will play a critical role in our efforts to find the specific superconformal field theory that is dual to superstrings moving on $AdS_3 \times S^3\times S^3 \times S^1$ geometry. In the present note we shall try to inch towards the root cause of the problem by analysing associated theories from a slightly different route by investigating the $N=4$ supersymmetries and $N=4$ superconformal symmetries relevant for the $AdS_3$ systems. Our premise is that the analysis of these structures will help us in pinpointing the specific superconformal field theory dual to superstrings moving on above geometry and settle this issue that is by now quarter of a century old.
 
In our recent analysis of Witten's original, our complementary and complete ADHM instanton sigma models and $AdS_3$ superstrings we discovered novel features of the supersymmetry structure as well as the structure of the corresponding moduli spaces.  In this note we take a stock of these features and give a revised and more comprehensive description of the emergent picture of this supersymmetry structure and the mappings between these three systems, namely, $N=4$ superconformal symmetries, supersymmetries of ADHM instanton sigma models and the superconformal symmetries of the conformal field theories dual to superstrings moving on $AdS_3 \times S^3\times K3$,  $AdS_3 \times S^3\times T^4$ and $AdS_3 \times S^3\times S^3 \times S^1$ geometries.

Plan of the rest of this note is as follows. First of all, in Section \ref{adhm}, we take up the discussion of the three ADHM instanton linear sigma models -  Witten's original model, Ali-Ilahi's complementary model and Ali-Salih's complete model. We focus upon  the corresponding $N=4$ supersymmetry structures. The theories described in this section have only supersymmetry and no conformal symmetry because of the finite size of the instanton solution that breaks the scale invariance and hence the conformal invariance. 

Next, in Section \ref{scas}, we collect information about $N=4$ superconformal algebras.  These are the symmetry algebras for two dimensional conformal field theories with four supercharges. There are three such symmetries. Corresponding to these there are three different $N=4$ superconformal algebras - small, middle and large. In this section we collect the operator content, corresponding operator product expansions as well as the relations between the three different algebras. Though major part of this information is well known and classic but we also include some additional information that is usually not explicitly specified. This additional information is necessary for specific applications in this note. At the end of this section we point out to which (super) conformal field theories the ADHM instanton linear sigma models flow in the infrared limit.

In Section \ref{ads3} we take up the identification of $N=4$ superconformal structures of the superconformal field theories dual to various $AdS_3$ superstrings moving in three different but specific back grounds - $AdS_3 \times S^3\times K3$,  $AdS_3 \times S^3\times T^4$ and $AdS_3 \times S^3\times S^3 \times S^1$ by pointing out that the dual superconformal field theories have small, middle and large $N=4$ superconformal symmetries. The consolidated and comprehensive map of these issues in our subjective view is very gratifying. In addition to that we have identified several issues to be dealt with.

We conclude in Section \ref{discussion}.

\section{ADHM  Sigma Model Supersymmetries}\label{adhm}

In this section we take up the task of recording the (0, 4) supersymmetries of ADHM instanton linear sigma models. Yang-Mills instantons are self-dual (or anti-self dual) solutions to Yang-Mills equations. These are field theory constructs. Sigma models enter into discussion in the process of generalization of above field theoretic solutions to string theory by incorporating gravitational, anti-symmetric tensor field and dilatonic backgrounds. 

The part objective of the present section  is to identify occurrences of $N=4$ supersymmetry structures in the context of ADHM instanton sigma models, namely, Witten's original\cite{Witten:1994tz}, Ali-Ilahi's complementary \cite{Ali:2023csc} and Ali-Salih's complete \cite{Ali:2023icn} ADHM instanton linear sigma models.

Instantons have a size and hence break conformal invariance. Because of this we get only the supersymmetry structure in these ADHM instanton sigma models and not the superconformal symmetry. It is believed that these theories flow to respective superconformal theories in the infrared limit. This is the limit where complete superconformal symmetry becomes operative in these systems.  

Our objective in this and next section is to map the $N=4$ supersymmetries of the ADHM instanton sigma models to appropriate $N=4$ superconformal symmetries. The same structure should be present in  case of 't Hooft instantons. The investigations in the case of the 't Hooft instantons were very thoroughly carried out in Refs. \cite{Strominger:1990et, Callan:1991dj, Callan:1991ky, Callan:1991at}.  The 't Hooft instanton is simpler in structure than the ADHM one. The former depends upon a restricted number of parameters while the latter realizes the full natural set of parameters. The stringy generalization of the ADHM instanton sigma was done in Ref.\cite{Witten:1994tz}. We call this the original ADHM instanton linear sigma model. In Ref.\cite{Ali:2023csc}(see also \cite{Ali:2023ams}) we constructed an ADHM instanton linear sigma model that we call the complementary model. We did yet another construction in Ref.\cite{Ali:2023icn} which we call the complete ADHM instanton linear sigma model. Meaning of this terminology should become apparent in the course of discussion. 

We shall take up the three models - original, complementary and the complete one, one at a time and point out corresponding supersymmetric structure in the rest of this section.

We now begin with  description of the action, field content of the original ADHM instanton sigma model in such a manner that the focus is upon the fact that it has an $N=4$ supersymmetry. We shall borrow the expression for action and notation from Ref.\cite{Lambert:1995dp}. The action is given below.

\begin{eqnarray}
 S &=& \int d^2 \sigma \biggl\{   \partial_{-}  {X}_{AY}\partial_{+} X^{AY}+ i\epsilon_{A'B'}\epsilon_{YZ}{\psi}_{-}^{A^{\prime}Y} \partial_{+} \psi_-^{B^{\prime}Z} \nonumber\\
&+&\partial_{-}{\phi}_{A^{\prime}Y^{\prime}}\partial_{+} \phi^{A^{\prime}Y^{\prime}}
+i\epsilon_{AB}\epsilon_{Y^{\prime}Z^{\prime}}\chi_-^{AY^{\prime}}\partial_{+} \chi_{-}^{BZ^{\prime}} 
+ i\lambda_+^a \partial_{-}\lambda_+^a \nonumber\\
&-& \frac{i}{2}m\lambda_+^a \Big ( \epsilon^{AB}\frac{\partial C^{a}_{AB'}}{\partial X^{BY}} \psi_-^{B'Y} 
+ \epsilon^{A'B'} \frac{\partial C^{a}_{AA'}}{\partial \phi^{B'Y'}} \chi_-^{AY'}\Big ) \nonumber \\
&-& \frac{1}{8}m^{2}\epsilon^{AB}\epsilon^{A'B'}C^a_{AA'}C^a_{BB'} \biggr\}.
   \label{action1}
\end{eqnarray}
The two dimensional light-come coordinates are 
\begin{equation}
\sigma^{\pm}=\frac{1}{2}(\tau \pm \sigma)
\end{equation}
with the following corresponding derivatives
\begin{equation}
\partial_{\pm} = \frac{1}{\sqrt{2}}(\partial_0 \pm \partial_1).
\end{equation}
where $\tau$ and $\sigma$ are the two dimensional worldsheet  coordinates. The worldsheet metric is
\begin{equation}
ds^2=d\tau^2-d\sigma^2.
\end{equation} 

The other indices are as follows.

The $4k$ bosons $X^{AY}$ with $A=1,2$ and $Y=1,2,...,2k$   are part of the standard multiplet and these have right handed superpartners $\psi_{-}^{A'Y}, A'=1,2$. We also have another set of $4k'$ bosons, $\phi^{A'Y'}$, part of the twisted multiplet, with corresponding superpartners $\chi_{-}^{AY'}$ with $Y'=1,2,...,2k'$. The raising and lowering of 
$Y, Z,...$ and $Y', Z' ,...$ indices is done with respective $Sp(k)$ and
$Sp(k')$ tensors $\epsilon^{YZ} (\epsilon_{YZ})$, $\epsilon^{Y'Z'} (\epsilon_{Y'Z'})$. Similarly the $A, B, ...$ and $A', B', ...$ indices  are
raised (lowered) with the help of anti-symmetric tensors $\epsilon^{AB} (\epsilon_{AB})$, $\epsilon^{A'B'} (\epsilon_{A'B'})$ respectively. The free theory has an $F \times F' \times H \times H'$ symmetry that acts on $AB,  A'B', YZ $ and $Y'Z'$ indices respectively where $F=SU(2)$, $F'=SU(2)$, $H=Sp(k)$ and $H'=Sp(k')$. This is generally broken by the potential terms. 

To begin with there is a $Z_2$ symmetry between the standard and the twisted multiplet with respect to the interchange of $F$ and $F'$ (and $H$ and $H'$). In the original construction this was broken because only the   $F'$ symmetry was retained. As a result we got an action for $k'$ instanton number and $4k+4k'$ target space dimensions. 

This $Z_2$ symmetry will also be broken in the upcoming complementary construction too because we shall retain only the $F$ symmetry and get the action for $k$ instanton number and $4k+4k'$ target space dimensions.

\subsection{Supersymmetry of Witten's Original Model}

The chiral sigma model (\ref{action1}) has $(0,4)$ on-shell supersymmetry
\begin{eqnarray}\label{susy}
\delta_{\eta} X^{AY}&=&i \epsilon_{A'B'}\eta^{AA'}_{+}\psi^{B'Y}_{-},
\delta_{\eta} \psi^{A'Y}_{-}= \epsilon_{AB}\eta^{AA'}_{+} \partial_{-} X^{BY}, \nonumber\\ 
\delta_{\eta} \phi^{A'Y'}&=&i \epsilon_{AB}\eta^{AA'}_{+}{\chi}^{BY'}_{-}, 
\delta_{\eta}{\chi}^{AY'}_{-} = \epsilon_{A'B'}\eta^{AA'}_{+} \partial_{-} \phi^{B'Y'}
\end{eqnarray}

There are supercharges $Q^{AA'}$ that obey the superalgebra 
\begin{equation}
\{Q^{AA'},Q^{BB'}\}=\epsilon^{AB}\epsilon^{A'B'}
P^+,~~~P^+=P_-=-i\partial/\partial \sigma^-.
\label{susyalgebra1}
\end{equation}
 The condition for $N=4$ supersymmetry is
\begin{eqnarray}
\label{adhmcondc}
  \frac{\partial C_{AA'}^a}{\partial X^{B Y}}
+\frac{\partial C_{BA'}^a}{\partial X^{A Y}}
=  \frac{\partial C_{AA'}^a}{\partial \phi^{B'Y'}}
+\frac{\partial C_{AB'}^a}{\partial \phi^{A'Y'}}=0.
\label{susycondition1}
\end{eqnarray}
To obtain explicit form of $C_{AA'}^a$ we begin with the following general form that is separately linear in both $\phi$ and $X$.
\begin{eqnarray}\label{witten2.22}
C_{AA'}^a&=M_{AA'}^a+X_{AY}N^{a\,Y}_{A'}
+\phi_{A'}{}^{Y'}D_{AY'}^a +X_{A}{}^{Y}\phi_{A'}{}^{Y'}E^a_{YY'}.
\end{eqnarray}

The condition for the Lagrangian to be invariant under all supersymmetries is
\begin{equation}
\label{caacond}
\sum_a\left(C_{AA'}^aC_{BB'}^a+C_{BA'}^aC_{AB'}^a
\right)=0.
\end{equation}

In Witten's original model the choice is to take $M=N=0$ such that 
\begin{eqnarray}\label{witten2.23}
{C}^{a}_{AA'}=\epsilon_{A'B'}D^{a}_{AY'}\phi^{~B'Y'}+ \epsilon_{AB}\epsilon_{A'B'}E^{a}_{YY'}X^{BY}\phi^{~B'Y'} \equiv \phi_{A'}^{~Y'}B^{a}_{AY'}(X)
\end{eqnarray}
such that the condition (\ref{caacond}) reduces to

\begin{equation}\label{B}
\sum_{a}(B^{a}_{AY'}B^{a}_{BZ'}+B^{a}_{BY'}B^{a}_{AZ'})=0.
\end{equation}
Here $B^{a}_{AY'}(X)$ is linear in $X$ and independent of $\phi$.

The resulting Yukawa and potential terms are
\begin{eqnarray}
 S' &=& \int d^2 \sigma \biggl\{ - \frac{i}{2}m\lambda_+^a \Big ( \epsilon^{AB}\frac{\partial C^{a}_{AB'}}{\partial X^{BY}} \psi_-^{B'Y} 
+ \epsilon^{A'B'} \frac{\partial C^{a}_{AA'}}{\partial \phi^{B'Y'}} \chi_-^{AY'}\Big ) \nonumber \\
&-& \frac{1}{8}m^{2}\epsilon^{AB}\epsilon^{A'B'}C^a_{AA'}C^a_{BB'} \biggr\}.
   \label{action2}
\end{eqnarray}
where $C^{a}_{AA'}$ is given by Eqn.(\ref{witten2.23}).

\subsection{Supersymmetry of the Complementary Model}

We now take up the description of the complementary ADHM instanton sigma model constructed by us in Ref.\cite{Ali:2023csc} (see also \cite{Ali:2023ams}). Once again our focus will be the fact that this model has an $N=4$ supersymmetry. Additionally we shall point out that the original and the complementary models are dual to each other and hence the $N=4$ supersymmetries of the two models are dual to each other.  

The $N=4$ supersymmetry in Witten's construction uses an $SU(2)$ symmetry $F'$. Apart from $F'$ there is another $SU(2)$ symmetry in the initial field content which is termed $F$. Witten had suggested that an alternative model using this should be constructed. This is what we carried out in Ref.\cite{Ali:2023csc, Ali:2023ams}. We call the resulting model as the complementary ADHM instanton sigma model. We believe that this too flows in the infrared to a small $N=4$ super conformal field theory. There is a duality between the complementary and the original models. This is the duality between the two $S(2)$ symmetries $F$ and $F'$.

The complementary model has $\hat M=\hat D=0$ in the following general form of the tensor $\hat{C}^{\hat a}_{AA'}$
\begin{eqnarray}\label{genchat}
\hat C_{AA'}^{\hat{a}}&=\hat M_{AA'}^{\hat{a}}+X_{AY}\hat N^{\hat a\,Y}_{A'}
+\phi_{A'}{}^{Y'}\hat D_{AY'}^{\hat{a}} +X_{A}{}^{Y}\phi_{A'}{}^{Y'}\hat E^{\hat{a}}_{YY'}.
\end{eqnarray}
results in
\begin{eqnarray}\label{caphi}
\hat{C}^{\hat a}_{AA'}=\epsilon_{AB} N^{\hat a}_{A'Y}X^{BY}+ \epsilon_{AB}\epsilon_{A'B'}E^{\hat a}_{YY'}X^{BY}\phi^{~B'Y'}\equiv X_{A}^{~Y}A^{\hat a}_{A'Y}(\phi).
\end{eqnarray}
The condition corresponding to Eqn.(\ref{caacond})
\begin{equation}
\sum_{\hat{a}}\left(\hat C_{AA'}^{\hat{a}}\hat C_{BB'}^{\hat{a}}+\hat C_{BA'}^{\hat{a}} \hat C_{AB'}^{\hat{a}}
\right)=0
\end{equation}
which becomes
\begin{equation}\label{aaacond}
\sum_{\hat{a}}(A^{\hat a}_{A'Y}A^{\hat a}_{B'Z}+A^{\hat a}_{B'Y}A^{\hat a}_{A'Z})=0.
\end{equation}
 Here $A^{\hat a}_{A'Y}(\phi)$ is linear in $\phi$ and independent of $X$.

 The resulting Yukawa and potential terms are
\begin{eqnarray}
 S'' &=& \int d^2 \sigma \biggl\{ - \frac{i}{2}m\lambda_+^a \Big ( \epsilon^{AB}\frac{\partial {\hat C}^{a}_{AB'}}{\partial X^{BY}} \psi_-^{B'Y} 
+ \epsilon^{A'B'} \frac{\partial {\hat C}^{a}_{AA'}}{\partial \phi^{B'Y'}} \chi_-^{AY'}\Big ) \nonumber \\
&-& \frac{1}{8}m^{2}\epsilon^{AB}\epsilon^{A'B'}{\hat C}^a_{AA'}{\hat C}^a_{BB'} \biggr\}.
   \label{action3}
\end{eqnarray}
where $\hat C^{a}_{AA'}$ is given by Eqn.(\ref{caphi}).

Both of the models, original and complementary, have (0, 4) supersymmetries but the two supersymmetries are different from each other. These two (0, 4) supersymmetries are related to each other by the simple duality: $X\leftrightarrow\phi$, $k\leftrightarrow k'$, $F\leftrightarrow F'$, $H\leftrightarrow H'$.

\subsection{Supersymmetry of the Complete Model}

We now summarize the complete ADHM instanton sigma model constructed by us in Ref.\cite{Ali:2023icn}. The Yukawa couplings in this model incorporate features of both of the $N=4$ supersymmetries. In this case
\begin{equation}
    S'''=S'+S''.\label{action4}
\end{equation}
As a result the supersymmetry algebra, though still an $N=4$ algebra, must be bigger than the earlier algebras. Explicit structure of this algebra has to be worked out. We do know a few glimpses of its structure. Whereas the supersymmetry algebras of the original and complementary algebras have on one $SU(2)$ R-symmetry each this one has $SU(2)\times SU(2)$. 

The duality between the original and complementary ADHM instanton linear sigma models now becomes a duality symmetry of the complete model. As a result there is no moduli space of the complete model.

\section{\texorpdfstring{$N=4$}{} Superconformal Algebras}\label{scas}

In this section we collect information about large, small and middle $N=4$ superconformal algebras. The information includes the field content, the operator product expansions and the inter-relations between these symmetries. Major part of this information is well known and classic but explicit expressions of some of the details are not available in the literature and we fill in the gaps in this section. The information collected in this section serves as the calibration standard for comparison with both previous as well as the next section.

First subsection is about large $N=4$ superconformal algebra, its field content and corresponding operator product expansions as well as features of its internal structure. These features are at the core of the insights in this note. This happens to be the largest two dimensional superconformal algebra with a central extension and very rich internal structure. In our view its internal $Z_2$ symmetry between the two $SU(2)$ Kac-Moody sub-algebras should play a role in settling the issue of completely specifying the $N=4$ superconformal field theory dual to superstrings moving in $AdS_3 \times S^3\times S^3 \times S^1$  background geometry\cite{Ali:2023kkf}. 

Next two subsections collect information about the small $N=4$ superconformal algebra. These subsections  also spells out how the small superconformal algebra sits inside the large one as a sub-algebra in two different ways related by a duality. Next two subsections collects the similar information about the middle $N=4$ superconformal algebra and how to obtain it from the large one by In\"on\"u-Wigner contraction in two different ways again related by a simple duality.

Final subsection of this section is about how the (0, 4) supersymmetries of Witten's original, Ali-Ilahi's complementary and Ali-Salih's complete ADHM instanton sigma models flow to respective $N=4$ superconformal symmetries summarized in the present section.

\subsection{The Large \texorpdfstring{$N=4$}{} Superconformal Algebra}

The large $N=4$ superconformal algebra has sixteen generators: $T(z)$, $G^a(z)$, $A^{\pm i}(z)$,$Q^a(z)$ and $U(z)$, $a=0,1,2,3$ and $i=1,2,3$. Currents $A^{\pm i}(z)$ generate two $SU(2)$ Kac-Moody algebras and $U(z)$ a $U(1)$ algebra. For the sake of completeness we list the corresponding operator product expansions below.

\begin{eqnarray}
T(z)T(z') & = &\frac{c/2}{(z-z')^4}+\frac{2T(z')}{(z-z')^2}+\frac{\partial T(z')}{z-z'}+\cdots,\nonumber\\
T(z){\cal O}(z') & = &\frac{d_{\cal O}{\cal O}(z')}{(z-z')^2}+\frac{\partial {\cal O}(z')}{z-z'}+\cdots,\nonumber\\
{\cal O} &\in& \{G^a, A^{\pm i}, Q^a, U\},\nonumber\\
d_{\cal O} &\in& \{3/2, 1, 1/2, 1\}\; {\rm respectively},\nonumber\\
G^a(z)G^b(z') & = &\frac{2c/3\delta^{ab}}{(z-z')^3}+\frac{8[\gamma\alpha^{+i}_{ab}A^{+i}(z')+(1-\gamma)\alpha^{-i}_{ab}A^{-i}(z')]}{(z-z')^2}\nonumber\\
&+&\frac{\{2T(z')\delta^{ab}-4[\gamma\alpha^{+i}_{ab}\partial A^{+i}(z')+(1-\gamma)\alpha^{-i}_{ab}\partial A^{-i}(z')]\}}{z-z'}+\cdots,\nonumber\\
A^{+i}(z)G^a(z') & = &\alpha^{+i}_{ab}\left[\frac{G^b(z')}{(z-z')}-\frac{2(1-\gamma)Q^b(z')}{(z-z')^2}\right]+\cdots,\nonumber\\
A^{\pm i}(z)A^{\pm j}(z') & = &-\frac{k^{\pm}/2\delta^{ij}}{(z-z')^2}+\frac{\epsilon^{ijk}A^{\pm k}(z')}{(z-z')}+\cdots,\nonumber\\
A^{+i}(z)A^{-j}(z') & = & A^{\pm i}(z)U(z') = 0,\nonumber\\
Q^a(z)G^b(z') & =&\frac{2[\alpha^{+i}_{ab}A^{+i}(z')-\alpha^-_{ab}A^{-i}(z')]}{(z-z')}+\cdots,\nonumber\\
A^{\pm i}(z)Q^a(z') & = & \frac{\alpha^{\pm i}_{ab}Q^b(z')}{z-z'} +\cdots; U(z)Q^a(z')=0,\nonumber
\end{eqnarray}
\begin{eqnarray}
U(z)G^a(z')&=&\frac{Q^a(z')}{z-z'}+\cdots, Q^a(z)Q^b(z') = -\frac{c}{12\gamma(1-\gamma)}\frac{\delta^{ab}}{z-z'}+\cdots,\nonumber\\
U(z)U(z')&=& -\frac{c}{12\gamma(1-\gamma)}\frac{1}{(z-z')^2}+\cdots.
\label{large1}   
\end{eqnarray}

The large $N=4$ superalgebra has two parameters which can be taken to be either $k^\pm$ or $c$ and $\gamma$. The relations between these are:
\begin{eqnarray}
    c&=&6k^+k^-/(k^++k^-),\nonumber\\ \gamma&=&k^-/(k^++k^-),\nonumber\\
    1-\gamma&=&k^+/(k^++k^-),\nonumber\\
    c/\gamma&=&6k^+,\nonumber\\
    c/(1-\gamma)&=&6k^-.
\end{eqnarray}

This superalgebra has a $Z_2$ symmetry under the following exchanges: $k^+\longleftrightarrow k^-$, $A^{+i}\longleftrightarrow A^{-i}$, $\alpha^{+i}_{ab}\longleftrightarrow \alpha^{-i}_{ab}$, $\gamma\longleftrightarrow 1-\gamma$. In our view this simple feature should play a significant role in settling the issue that served as a motivation for collecting the information in this note, that is, to find the specific superconformal field theory dual to superstrings moving on $AdS_3 \times S^3\times S^3 \times S^1$ manifold.

\subsection{The Small \texorpdfstring{$N=4$}{} Superconformal Algebra : I}

The large $N=4$ superalgebra contains the small $N=4$ superconformal algebras as a sub-algebra in two different ways. First one is obtained using the following redefinitions:
\begin{eqnarray}
  {\tilde T}(z) &=& T(z) - (1-\gamma)\partial U(z),\nonumber\\
  {\tilde G}^a(z) &=& G^a(z) - 2(1-\gamma)\partial Q^a(z).
\end{eqnarray}

This small $N=4$ superconformal sub-algebra has eight 
generators: ${\tilde T}(z)$, ${\tilde G}^a(z)$, $A^{+i}(z)$. The other fields drop out of the operator product expansions. The central charge of this sub-algebra is
\begin{equation}
    {\tilde c}=c/6\gamma.
\end{equation}
The level of the $SU(2)$ Kac-Moody sub-algebra is given by 
\begin{equation}
    {\tilde k}=c/6.
\end{equation}

Corresponding operator product expansions are given below where we have removed the tilde from various quantities to avoid cluttering of notation.

\begin{eqnarray}
T(z)T(z') & = &\frac{c/2}{(z-z')^4}+\frac{2T(z')}{(z-z')^2}+\frac{\partial T(z')}{z-z'}+\cdots,\nonumber\\
T(z){G^a}(z') & = &\frac{3/2{G^a}(z')}{(z-z')^2}+\frac{\partial G^a(z')}{z-z'}+\cdots,\nonumber\\
T(z){A^{+i}}(z') & = &\frac{{A^{+i}}(z')}{(z-z')^2}+\frac{\partial A^{+i}(z')}{z-z'}+\cdots,\nonumber\\
G^a(z)G^b(z') & = &\frac{2c/3\delta^{ab}}{(z-z')^3}+\frac{8\alpha^{+i}_{ab}A^{+i}(z')}{(z-z')^2}+\frac{\{2T(z')\delta^{ab}-\alpha^{+i}_{ab}\partial A^{+i}(z')\}}{z-z'}+\cdots,\nonumber\\
A^{+i}(z)G^a(z') & = &\frac{\alpha^{+i}_{ab}G^b(z')}{(z-z')}+ \cdots,\nonumber\\
A^{+i}(z)A^{+j}(z') & = &-\frac{k/2\delta^{ij}}{(z-z')^2}+\frac{\epsilon^{ijk}A^{+k}(z')}{(z-z')}+\cdots.
\label{small1}   
\end{eqnarray}

Eight of the operators, namely, $A^{-i}(z), Q^a(z)$ and $U(z)$ drop out of these OPEs. These can be cast into a hypermultiplet of small $N=4$ superconformal algebra. The OPEs of the operators of the hypermultiplet with the operators of the small $N=4$ superconformal algebra are given below.

\begin{eqnarray}
T(z)A^{-i}(z') & = &\frac{A^{-i}(z')}{(z-z')^2}+\frac{\partial A^{-i}(z')}{z-z'}+\cdots,\nonumber\\
T(z)Q^a(z') & = &\frac{1/2Q^a(z')}{(z-z')^2}+\frac{\partial Q^a(z')}{z-z'}+\cdots,\nonumber\\
T(z)U(z') & = &\frac{2c}{\gamma}\frac{1}{(z-z')^3}+\frac{U(z')}{(z-z')^2}+\frac{\partial U(z')}{z-z'}+\cdots,\nonumber\\
A^{-i}(z)G^a(z') & = &\alpha^{-i}_{ab}\left[\frac{G^b(z')}{z-z'}+\frac{2Q^b(z')}{(z-z')^2}\right]+\cdots,\nonumber\\
Q^a(z)G^b(z') & = &\frac{c}{(z-z')^2}+\frac{2[\alpha^{+i}_{ab}A^{+i}(z')-\alpha^-_{ab}A^{-i}(z')]}{z-z'}+\cdots,\nonumber\\
U(z)G^a(z')&=&\frac{Q^a(z')}{z-z'}+\cdots,\nonumber\\
A^{+i}(z)A^{-j}(z')&=&0,\nonumber\\
A^{+i}(z)Q^a(z')&=&\frac{\alpha^{+i}_{ab}Q^b(z')}{z-z'}\nonumber\\
A^{+i}(z)U(z')&=&0.\label{hyper1}
\end{eqnarray}

The OPEs of the operators of the hypermultiplet among each other are given below.
\begin{eqnarray}
U(z)A^{-i}(z')&=&0,\nonumber\\
U(z)Q^a(z')&=&0,\nonumber\\
U(z)U(z')&=&-\frac{c}{12\gamma(1-\gamma)}\frac{1}{(z-z')^2},\nonumber\\
A^{-i}(z)Q^a(z')&=&\frac{\alpha^{-i}_{ab}Q^b(z')}{z-z'},\nonumber\\
Q^a(z)Q^b(z')&=& -\frac{c}{12\gamma(1-\gamma)}\frac{\delta^{ab}}{z-z'},\nonumber\\
Q^a(z)U(z')&=&0,\nonumber\\
A^{-i}(z)A^{-j}(z') & = &-\frac{k^-/2\delta^{ij}}{(z-z')^2}+\frac{\epsilon^{ijk}A^{-k}(z')}{z-z'}+\cdots.\label{hyper2}
\end{eqnarray}

Here two OPEs, $T(z)U(z')$ and $Q^a(z)G^b(z')$ have unexpected anomalous central terms. We shall not try to sort out this issue here and absorb this feature in the definition of the hypermultiplet.

Above two sets of operator product expansions, (\ref{hyper1}) and (\ref{hyper2}), are not available in literature to our best knowledge. 

\subsection{The Small \texorpdfstring{$N=4$}{} Superconformal Algebra : II}

The second small $N=4$ superconformal sub-algebra is obtained using following redefinitions:
\begin{eqnarray}
  {\hat T}(z) &=& T(z) - \gamma\partial U(z),\nonumber\\
  {\hat G}^a(z) &=& G^a(z) - 2\gamma\partial Q^a(z).
\end{eqnarray}
This small $N=4$ superconformal sub-algebra has eight 
generators: ${\hat T}(z)$, ${\hat G}^a(z)$, $A^{-i}(z)$. The other fields drop out of the operator product expansions. The central charge of this sub-algebra is
\begin{equation}
    {\hat c}=c/6(1-\gamma).
\end{equation}
This time the level of the $SU(2)$ Kac-Moody sub-algebra is
\begin{equation}
    {\hat k}=c/6.
\end{equation}
Corresponding operator product expansions are given below where once again we have removed the caret from various quantities to avoid cluttering of the notation.
\begin{eqnarray}
    T(z)T(z') & = &\frac{c/2}{(z-z')^4}+\frac{2T(z')}{(z-z')^2}+\frac{\partial T(z')}{z-z'}+\cdots,\nonumber\\
T(z){G^a}(z') & = &\frac{3/2{G^a}(z')}{(z-z')^2}+\frac{\partial G^a(z')}{z-z'}+\cdots,\nonumber\\
T(z){A^{-i}}(z') & = &\frac{3/2{A^{-i}}(z')}{(z-z')^2}+\frac{\partial A^{-i}(z')}{z-z'}+\cdots,\nonumber\\
G^a(z)G^b(z') & = &\frac{2c/3\delta^{ab}}{(z-z')^3}+\frac{8\alpha^{-i}_{ab}A^{-i}(z')}{(z-z')^2}+\frac{\{2T(z')\delta^{ab}-\alpha^{-i}_{ab}\partial A^{-i}(z')\}}{z-z'}+\cdots,\nonumber\\
A^{-i}(z)G^a(z') & = &\frac{\alpha^{-i}_{ab}G^b(z')}{z-z'}+ \cdots,\nonumber\\
A^{-i}(z)A^{-j}(z') & = &-\frac{k/2\delta^{ij}}{(z-z')^2}-\frac{\epsilon^{ijk}A^{\pm k}(z')}{z-z'}+\cdots.
\label{small2}   
\end{eqnarray}

This time the eight generators that drop out are $A^{+i}(z), Q^a(z)$ and $U(z)$. Once again these can be cast into a hyper-multiplet of small $N=4$ superconformal algebra. This time the OPEs of the operators of the hyper-multiplet with the operators of the small $N=4$ superconformal algebra are given below.

\begin{eqnarray}
T(z)A^{-i}(z') & = &\frac{A^{-i}(z')}{(z-z')^2}+\frac{\partial A^{-i}(z')}{z-z'}+\cdots,\nonumber\\
T(z)Q^a(z') & = &\frac{1/2Q^a(z')}{(z-z')^2}+\frac{\partial Q^a(z')}{z-z'}+\cdots,\nonumber\\
T(z)U(z') & = &\frac{2c}{1-\gamma}\frac{1}{(z-z')^3}+\frac{U(z')}{(z-z')^2}+\frac{\partial U(z')}{z-z'}+\cdots,\nonumber\\
A^{+i}(z)G^a(z') & = &\alpha^{+i}_{ab}\left[\frac{G^b(z')}{z-z'}+\frac{2Q^b(z')}{(z-z')^2}\right]+\cdots,\nonumber\\
Q^a(z)G^b(z') & = &\frac{c}{(z-z')^2}+\frac{2[\alpha^{+i}_{ab}A^{+i}(z')-\alpha^-_{ab}A^{-i}(z')]}{z-z'}+\cdots,\nonumber\\
U(z)G^a(z')&=&\frac{Q^a(z')}{z-z'}+\cdots,\nonumber\\
A^{+i}A^{-j}(z')&=&0,\nonumber\\
A^{-i}(z)Q^a(z')&=&\frac{\alpha^{-i}_{ab}Q^b(z')}{z-z'}\nonumber\\
A^{-i}(z)U(z')&=&0.\label{hyper3}
\end{eqnarray}

The operator product expansions of the operators of this hyper-multiplet among each other are given below.
\begin{eqnarray}
U(z)A^{+i}(z')&=&0,\nonumber\\
U(z)Q^a(z')&=&0,\nonumber\\
U(z)U(z')&=&-\frac{c}{12\gamma(1-\gamma)}\frac{1}{(z-z')^2},\nonumber\\
A^{+i}Q^a(z')&=&\frac{\alpha^{+i}_{ab}Q^b(z')}{z-z'},\nonumber\\
Q^a(z)Q^b(z')&=& -\frac{c}{12\gamma(1-\gamma)}\frac{\delta^{ab}}{z-z'},\nonumber\\
Q^a(z)U(z')&=&0,\nonumber\\
A^{+i}(z)A^{+j}(z') & = &-\frac{k^+/2\delta^{ij}}{(z-z')^2}+\frac{\epsilon^{ijk}A^{-k}(z')}{z-z'}+\cdots.\label{hyper4}
\end{eqnarray}

Once two OPEs, $T(z)U(z')$ and $Q^a(z)G^b(z')$ have unexpected anomalous central terms and once again we shall not try to sort out this issue here and absorb this feature in the definition of the hypermultiplet. Once again the last two sets of operator product expansions are not available in the literature.

These two embeddings of the small $N=4$ superconformal algebra in the large $N=4$ superconformal algebra are independent of each other but there is a duality between these embeddings. This duality is : $\gamma\longleftrightarrow 1-\gamma$, $A^{+i}(z)\longleftrightarrow A^{-i}(z)$ and  $\alpha^{+i}_{ab}\longleftrightarrow \alpha^{-i}_{ab}$ and $k^+\longleftrightarrow k^-$. 

It is obvious that this duality between two embeddings of the small $N=4$ superconformal algebra as a sub-algebra of the large $N=4$ superconformal algebra is a result of the $Z_2$ automorphism of the latter.

\subsection{Infrared Limit of ADHM Sigma Models}

In this section we map the $N=4$ supersymmetries of ADHM instanton linear sigma models onto the $N=4$ superconformal algebras. Various ADHM instanton linear sigma models flow to corresponding superconformal field theories in the infrared limit.

The small $N=4$ superconformal algebra is contained in the large $N=4$ superconformal algebra in two ways related by a duality\cite{SEVRIN1988447, Ivanov:1988rt, Schoutens:1988ig, Ali:2000zu}.

Witten's original ADHM instanton linear sigma model flows to one of these small $N=4$ superconformal algebras. Ali-Ilahi's complementary ADHM instanton linear sigma model flows to the other small $N=4$ superconformal algebra and there is duality between the two cases.

Ali-Salih's complete ADHM instanton linear sigma model flows in the infrared to the full large $N=4$ superconformal algebra.

This is the first part of our main result in this note.

\subsection{The Middle \texorpdfstring{$N=4$}{} Superconformal Algebra : I}

We now take up the middle $N=4$ superconformal algebra.

The middle $N=4$ superconformal algebra is obtained from the large one by In\"on\"u-Wigner contraction. There are two independent contractions that give us the middle $N=4$ superconformal algebra from the large one.

The middle $N=4$ superconformal algebra is not a sub-algebra of the large one. It is obtained by In\"on\"u-Wigner contraction of the large one in two different ways. Once again the two resulting middle $N=4$ superconformal algebras thus obtained are related to each other by a duality.

In the first case we make the following singular limits:
\begin{eqnarray}
{\tilde T}(z) &=& \lim_{\gamma\rightarrow 1}T(z), \; {\tilde U}^i(z) = \lim_{\gamma\rightarrow 1}\sqrt{1-\gamma}A^{-i}(z),\nonumber\\
  {\tilde G}^a(z) &=& \lim_{\gamma\rightarrow 1} G^a(z),\; {\tilde Q}^a(z) = \lim_{\gamma\rightarrow 1}\sqrt{1-\gamma}Q^a(z),\nonumber\\ 
  {\tilde A}^{+i}(z) &=& \lim_{\gamma\rightarrow 1} A^{+i}(z),\; {\tilde U}(z) = \lim_{\gamma\rightarrow 1}\sqrt{1-\gamma}U(z).
  \label{limits1}
\end{eqnarray}

The resulting OPEs of the middle $N=4$ superconformal algebra are given below. Here we have removed the tilde to avoid cluttering notation and taken $U^0(z)=U(z)$.

\begin{eqnarray}
 T(z)T(z') & = &\frac{c/2}{(z-z')^4}+\frac{2T(z')}{(z-z')^2}+\frac{\partial T(z')}{z-z'}+\cdots,\nonumber\\
T(z){\cal O}(z') & = &\frac{d_{\cal O}{\cal O}(z')}{(z-z')^2}+\frac{\partial {\cal O}(z')}{z-z'}+\cdots,\nonumber\\
{\cal O} &\in& \{G^a, A^{+i}, Q^a, U^a\},\nonumber\\
d_{\cal O} &\in& \{3/2, 1, 1/2, 1\}\; {\rm respectively},\nonumber\\
G^a(z)G^b(z') & = &\frac{2c/3\delta^{ab}}{(z-z')^3}+\frac{8\alpha^{+i}_{ab}A^{+i}(z')}{(z-z')^2}+\frac{\{2T(z')\delta^{ab}-\alpha^{+i}_{ab}\partial A^{+i}(z')\}}{z-z'}+\cdots,\nonumber\\
A^{+i}(z)G^a(z') & = &\frac{\alpha^{+i}_{ab}G^b(z')}{z-z'}+ \cdots, U^i(z)G^a(z')  = \frac{\alpha^{-i}_{ab}Q^b(z')}{(z-z')^2}+ \cdots,\nonumber\\
A^{+i}(z)A^{+j}(z') & = &-\frac{k^+/2\delta^{ij}}{(z-z')^2}-\frac{\epsilon^{ijk}A^{\pm k}(z')}{z-z'}+\cdots, \nonumber\\
U^a(z)U^b(z') & = & -\frac{k^+/2\delta^{ab}}{(z-z')^2} + \cdots, A^{+i}(z)U^a(z') = 0 = U^a(z)Q^b(z'),\nonumber
\end{eqnarray}
\begin{eqnarray}
U(z)G^a(z') &=& \frac{Q^a(z')}{(z-z')^2} + \cdots, Q^a(z)G^b(z') =\frac{\delta^{ab}U(z')-2\alpha^{-i}_{ab}U^i(z')}{z-z'} + \cdots,
\nonumber\\
A^{+i}(z)Q^a(z') &=& \frac{\alpha^{+i}_{ab}Q^b(z')}{z-z'} + \cdots, Q^a(z)Q^b(z') = -\frac{k^-/2\delta^{ab}}{z-z'} + \cdots.
\label{middle1}
\end{eqnarray}

\subsection{The Middle \texorpdfstring{$N=4$}{} Superconformal Algebra : II}

In the second case we make the following singular limits:
\begin{eqnarray}
 {\hat T}(z) &=& \lim_{\gamma\rightarrow 0}T(z),\;  {\bar U}^i(z) = \lim_{\gamma\rightarrow 0}\sqrt{\gamma}A^{+i}(z),\nonumber\\
  {\hat G}^a(z) &=& \lim_{\gamma\rightarrow 0} G^a(z),\; {\hat Q}^a(z) = \lim_{\gamma\rightarrow 0}\sqrt{\gamma}Q^a(z),\nonumber\\ 
  {\hat A}^{-i}(z) &=& \lim_{\gamma\rightarrow 0} A^{-i}(z),\; {\hat U}(z) = \lim_{\gamma\rightarrow 0}\sqrt{\gamma}U(z).
  \label{limits2}
\end{eqnarray}

The resulting OPEs of the second middle $N=4$ superconformal algebra are given below. Here once again we have removed the bar to avoid cluttering notation and taken $U^0(z)=U(z)$.

\begin{eqnarray}
 T(z)T(z') & = &\frac{c/2}{(z-z')^4}+\frac{2T(z')}{(z-z')^2}+\frac{\partial T(z')}{z-z'}+\cdots,\nonumber\\
T(z){\cal O}(z') & = &\frac{d_{\cal O}{\cal O}(z')}{(z-z')^2}+\frac{\partial {\cal O}(z')}{z-z'}+\cdots,\nonumber\\
{\cal O} &\in& \{G^a, A^{-i}, Q^a, U^a\},\nonumber\\
d_{\cal O} &\in& \{3/2, 1, 1/2, 1\}\; {\rm respectively},\nonumber\\G^a(z)G^b(z') & = &\frac{2c/3\delta^{ab}}{(z-z')^3}+\frac{8\alpha^{-i}_{ab}A^{-i}(z')}{(z-z')^2}+\frac{\{2T(z')\delta^{ab}-\alpha^{-i}_{ab}\partial A^{-i}(z')\}}{z-z'}+\cdots,\nonumber\\
A^{-i}(z)G^a(z') & = &\frac{\alpha^{-i}_{ab}G^b(z')}{z-z'}+ \cdots,U^i(z)G^a(z')  = \frac{\alpha^{-i}_{ab}Q^b(z')}{(z-z')^2}+ \cdots,\nonumber
\end{eqnarray}
\begin{eqnarray}
A^{-i}(z)A^{-j}(z') & = &-\frac{k/2\delta^{ij}}{(z-z')^2}-\frac{\epsilon^{ijk}A^{\pm k}(z')}{z-z'}+\cdots,\nonumber\\
U^a(z)U^b(z') & = & -\frac{k^+/2\delta^{ab}}{(z-z')^2} + \cdots, A^{+i}(z)U^a(z') = 0 = U^a(z)Q^b(z'),\nonumber\\
U(z)G^a(z') &=& \frac{Q^a(z')}{(z-z')^2} + \cdots, Q^a(z)G^b(z') =\frac{\delta^{ab}U(z')-2\alpha^{-i}_{ab}U^i(z')}{z-z'} + \cdots,
\nonumber\\
A^{+i}(z)Q^a(z') &=& \frac{\alpha^{+i}_{ab}Q^b(z')}{z-z'} + \cdots, Q^a(z)Q^b(z') = -\frac{k^-/2\delta^{ab}}{z-z'} + \cdots.
\label{middle2}
\end{eqnarray}

The novelty here, as compared to the existing literature, is that there is a duality between two contractions given by $\gamma\longleftrightarrow 1-\gamma$, $A^{+i}(z)\longleftrightarrow A^{-i}(z)$ and  $\alpha^{+i}_{ab}\longleftrightarrow \alpha^{-i}_{ab}$ and $k^+\longleftrightarrow k^-$. Here two this duality is a result of the $Z_2$ inner automorphism of the large $N=4$ superconformal algebra. We also would like to emphasize that the expressions for the second contraction are not available in literature.

\subsection{The Mapping So Far}

The objective of this note is to spell out the mapping between ADHM instanton linear sigma models, $N=4$ superconformal algebras and the Type IIB superstrings moving on $AdS_3 \times S^3\times K3$,  $AdS_3 \times S^3\times T^4$ and $AdS_3 \times S^3\times S^3 \times S^1$ geometries. We have so far summarized the ADHM instanton linear sigma models and $N=4$ superconformal algebras. With this much of data parts of the intended mapping is already in place. This is explained below.

Witten's original and Ali-Ilahi's complementary ADHM instanton linear sigma models flow to the two embeddings of the small $N=4$ superconformal algebra in the large $N=4$ superconformal algebra. Before the RG flow the supersymmetries of the original and complementary ADHM models flow map onto the two small $N=4$ embeddings in large $N=4$ superconformal algebra. The duality between the two ADHM instanton sigma models is mapped onto the duality between the two small $N=4$ superconformal algebras. 

The larger $N=4$ supersymmetry of Ali-Salih's complete ADHM instanton linear sigma model flows in the infrared to the large $N=4$ superconformal algebra. Before the renormalization group flow the supersymmetry of the complete ADHM instanton linear sigma model is mapped onto the large $N=4$ superconformal algebra. This mapping is apparent in spite of the fact that large $N=4$ supersymmetry of the complete ADHM model has not been worked out in detail. 

To bring the middle $N=4$ superconformal algebra into picture we need a discussion of the $AdS_3$ superstrings.

\section{\texorpdfstring{$AdS_3$}{} Superstrings}\label{ads3}

All the $N=4$ superconformal symmetries are realized in case of the $AdS_3/CFT_2$ correspondence in their full glory. A symmetry is an abstract group structure. A realization offers an opportunity to see that symmetry in a concrete form. To encounter a symmetry in an actual physical system is even more gratifying. That is what happens in case of $AdS_3$ superstrings in the context of $N=4$ superconformal symmetries.

We now begin the analysis of the $N=4$ superconformal symmetries of the conformal field theories dual to superstrings moving on $AdS_3$ backgrounds. A lot is already known about this issue but the map gets substantially revised in view of our findings.

We shall first take up the results about two dimensional superconformal field theory dual to superstrings moving on $AdS_3\times S^3\times S^3\times S^1$ background geometry in Section \ref{ads3331}. It is well known that the corresponding superconformal algebra is the large $N=4$. Present investigations has been inspired by the fact that the specific superconformal field theory dual to above superstrings is not known precisely in general case. We point out our investigations that should help in solving this long standing problem.

Next, in Section \ref{ads333}, we take up the superstrings moving in the background geometry $AdS_3\times S^3\times K3$ and the two dimensional superconformal field theory dual to it. It is well known that the corresponding superconformal algebra is the small $N=4$ one. In Ref.\cite{Ali:2023xov} we discovered that there are two versions of these superstrings. These two superstrings, as well as the corresponding dual superconformal field theories, are related to each other by duality.

Next we take up the two dimensional superconformal field theory dual to superstrings moving on $AdS_3\times S^3\times T^4$ background geometry in Section \ref{ads334}. Corresponding dual superconformal field theory should have a middle $N=4$ superconformal algebra as the symmetry algebra. This algebra is obtained by contraction of the large $N=4$ superconformal algebra\cite{Ali:1993sd, Hasiewicz:1989vp, Ali:2003aa}. After our earlier investigations it becomes apparent that this process of contraction too can be carried out in two different ways that are related to each other by a simple duality.

We conclude this section with a summary of the structure that we have uncovered in Section \ref{sum}.

In this section we shall take advantage of the Giveon-Kutasov-Seiberg construction \cite{Giveon:1998ns} that gives  us a concrete method to construct the superconformal field theory relevant for $AdS_3$ superstrings.

\subsection{\texorpdfstring{$AdS_3\times S^3\times S^3\times S^3\times S^1$}{} Superstrings}\label{ads3331}

The case Type IIB superstrings moving in $AdS_3\times S^3\times S^3\times S^3\times S^1$ background is the main driving force behind present investigations. The reason is that we know a lot about the dual superconformal field theory to above superstrings under the AdS/CFT correspondence \cite{Elitzur:1998mm, deBoer:1999gea, Gukov:2004ym, Tong:2014yna, Eberhardt:2017fsi, Eberhardt:2017pty, Papadopoulos:2024uvi, Witten:2024yod} but we do not know it completely.

$AdS_3$ superstrings, in general, are defined in terms of D-brane configurations that go over to the requisite geometry in the near horizon limit. In this note we are interested mainly in three geometries - $AdS_3\times S^3\times S^3\times S^3\times S^1$, $AdS_3\times S^3\times S^3\times K3$ and $AdS_3\times S^3\times S^3\times T^4.$ These are obtained using three different D-brane configuration. In fact requisite D-brane configurations only partially known. Most well known case is of  $AdS_3\times S^3\times S^3\times K3$ which is realized as the Higgs branch of coincident D1-D5 branes. Yet the tragedy in this case is the zero binding energy of the D-branes\cite{Seiberg:1999xz}. The $AdS_3\times S^3\times S^3\times T^4$ superstrings have the additional issue of four $U(1)$'s. In fact this geometry can be obtained by Penrose limit of the $AdS_3\times S^3\times S^3\times S^3\times S^1$ geometry. Unfortunately we do not know the brane configuration that goes over to the latter geometry in the near horizon limit. The closest we can reach was suggested in Ref.\cite{Tong:2014yna}. He suggested a configuration of $Q_1$ D1-branes, $Q_5^+$ D5 branes and $Q_5^-$ D5$'$ branes. 
\begin{center}
    $Q_5^+$ D5-Branes : 012345\nonumber\\
    $Q_5^-$ D5$'$-Branes : 0****56789\nonumber\\
    $Q_5^+$ D5-Branes : 0****5\nonumber\\
\end{center}
The resulting configuration gives $AdS_3\times S^3\times S^3\times S^3\times R$ geometry in the near horizon limit. In the present note we shall not be able to improve upon this state of affairs.

Witten also suggested a construction of a model that uses both of the $SU(2)$ symmetries $F$ and $F'$. We did corresponding construction in Ref.\cite{Ali:2023icn}.  Since this model has two $SU(2)$ symmetries it should flow in the infrared to a large $N=4$ super conformal field theory.   It is already well known that the superconformal field theory dual to Type IIB superstrings moving  on  $AdS_3\times S^3\times S^3\times S^1$ backgrounds is the large $N=4$ SCFT. The duality between $F$ and $F'$ of the original and complementary models now becomes a $Z_2$ symmetry between them. This is the $Z_2$ symmetry between the two $S^3$'s of the $AdS_3\times S_+^3\times S_-^3\times S^1$ background, that is,  $S^3_+\leftrightarrow S^3_-$.

Now we argue how to get the large $N=4$ superconformal symmetry in this case. In  Giveon-Kutasov-Seiberg construction the isometry associated with the $SL(2, R)$ symmetry of $AdS_3$ gives rise to the Virasoro algebra. The isometry of the three sphere $S^3_+$ gives rise to one $SU(2)_+$ Kac-Moody symmetry while the other three sphere $S^3_-$ gives rise to one $SU(2)_-$ Kac-Moody symmetry. The lone $S^1$ gives rise to the solo $U(1)$ Kac-Moody symmetry. The large $N=4$ superconformal algebra is the only one that accommodates this much of Kac-Moody symmetry. Hence the $N=4$ superconformal symmetry relevant in case of Type IIB superstrings moving in $AdS_3\times S_+^3\times S_-^3\times S^1$ background is the large one.

We shall keep the details of the construction to a future note\cite{Ali:2024ca}.

\subsection{\texorpdfstring{$AdS_3\times S^3\times K3$}{} Superstrings}\label{ads333}

In his seminal paper on AdS/CFT correspondence Maldacena defined the $AdS_3\times S^3\times K3$ superstrings as the Higgs branch of the near horizon limit of $Q_1$ D1 branes and $Q_5$ D5 branes. We have already mentioned the issue of zero binding energy of this configuration above\cite{Seiberg:1999xz}. In this note we shall not say anything about this issue.

The task in this section is to construct the superconformal symmetry with $N=4$ that is relevant for Type IIB superstrings moving on $AdS_3\times S^3\times K3$ background. This is the case that Giveon-Kutasov-Seiberg dealt with (see also \cite{Eberhardt:2019qcl}). In this case too the Virasoro algebra is derived from the $SL(2, R)$ isometry of the $AdS_3$ space and the solo $SU(2)$ Kac-Moody algebra is derived from the isometry of the single three sphere $S^3$. There is no other Kac-Moody symmetry because $K3$ possesses no isometry. The $N=4$ algebra having above Kac-Moody symmetry is the small one. If we motivate the symmetry structure from $AdS_3\times S_+^3\times S_-^3\times S^1$ geometry then we get two ways to do so because we might retain either of the two three spheres. Once again we shall deal with the details of the construction in another note\cite{Ali:2024cb}.

\subsection{\texorpdfstring{$AdS_3\times S^3\times T^4$}{} Superstrings}\label{ads334}

We now take up the case of Type IIB superstrings moving on $AdS_3\times S^3\times T^4$ geometry and the superconformal symmetry of the dual field theory.

Following Maldacena in his seminal work on AdS/CFT correspondence the $AdS_3\times S^3\times T^4$ superstrings, like $AdS_3\times S^3\times K3$ superstrings,  are defined as the Higgs branch of the near horizon limit of $Q_1$ D1 branes and $Q_5$ D5 branes. The issue of four torus is clearly in addition to above ansatz.

Corresponding superconformal field theory has the middle $N=4$ superconformal as the symmetry algebra.

It is the missing piece in the ADHM instanton linear sigma models. Luckily it is readily available in the other corner of the map we are drawing in this note. It is obtained by In\"on\"u-Wigner contraction of the large $N=4$ super conformal algebra \cite{Ali:1993sd, Hasiewicz:1989vp, Ali:2003aa}. From the insights that we gained from the structure of the ADHM instanton sigma models we realize that this contraction can be done in two ways and we get two middle $N=4$ super conformal algebras related to each other by a duality. In case of $AdS_3$ superstrings we get $AdS_3\times S_+^3\times T^4$ and $AdS_3\times S_-^3\times T^4$ by two different Penrose limits  of $AdS_3\times S_+^3\times S_-^3\times S^1$ of background \cite{Sommovigo:2003kd, Dei:2018yth, Ali:2024cc}.

In this case the superconformal field theory has the energy-momentum tensor $T(z)$ and the Virasoro algebra is realized using the $SL(2, R)$ isometry of $AdS_3$ part of the geometry. There are four supercurrents $G^a(z), a=0,1,2,3$, just like the other two cases. Isometries of the three sphere $S^3$ gives rise to the internal Kac-Moody algebra of the middle $N=4$ superconformal algebra. There are four spin half fermionic current $Q^a(z)$. Finally there are four U(1) currents $U^\alpha, \alpha=0,1,2,3.$

\subsection{Summing Up}\label{sum}

We shall now update our earlier assessment about the structures that we have uncovered in this note. There are abstract symmetries with $N=4$ superconformal groups and corresponding to small, middle and large $N=4$ superconformal algebras. Groups and algebras have representations and realizations in terms of concrete mathematical entities. Corresponding things in case of the superconformal field theories, for example, are the free field realizations. Finally there are real physical systems possessing such symmetries. $AdS_3$ superstrings are our second such system, ADHM instanton sigma models being the first one.

With the data collected in above two sections the following world map of ADHM instanton sigma models, $N=4$ superconformal algebras and $AdS_3$ superstrings crystallizes. 

The $N=4$ supersymmetry of Witten's ADHM instanton linear sigma model is mapped onto one of the embeddings of the small $N=4$ superconformal algebra inside the large $N=4$ superconformal algebra. This in turn happens to the superconformal symmetry of the Type IIB superstrings moving on a manifold with $AdS_ 3 \times S^3 \times K3$ geometry. The $N=4$ supersymmetry of Ali-Ilahi's ADHM instanton linear sigma model is mapped onto the other embedding of the small $N=4$ superconformal algebra inside the large $N=4$ superconformal algebra. This in turn is mapped onto the other formalism of the Type IIB superstrings moving on a manifold with $AdS_ 3 \times S^3 \times K3$ geometry. 

First one of these mappings can be represented as follows. 
\begin{center}
Original ADHM Sigma Model

$\downarrow$ 

(Small $N=4^+$ SCFT)

$\downarrow$

 $AdS_ 3 \times S_+^3 \times K3$
\end{center}

Second mapping can be represented as follows.
\begin{center}
Complementary ADHM Sigma Model

$\downarrow$ 

(Small $N=4^-$ SCFT)

$\downarrow$

 $AdS_ 3 \times S_-^3 \times K3$
\end{center}

The superscript + on $N=4^+$ and the subscript + on $S^3_+$ represent the fact that we are using one embedding of small $N=4$ in the large $N=4$.

Similarly the superscript - on $N=4^-$ and the subscript - on $S^3_-$ represent the fact that we are using the other embedding of small $N=4$ in the large $N=4$.

Two middle $N=4$ superconformal algebras are mapped onto two Type IIB superstrings moving on a manifold with $AdS_ 3 \times S^3 \times T^4$ geometry.

First one of these mappings can be represented as follows. 
\begin{center}
(Middle $N=4^+$ SCFT)

$\downarrow$

 $AdS_ 3 \times S_+^3 \times T^4$
\end{center}

Second mapping can be represented as follows.
\begin{center}
(Middle $N=4^-$ SCFT)

$\downarrow$

 $AdS_ 3 \times S_-^3 \times T^4$
\end{center}

Once again the superscript + on $N=4^+$ and the subscript + on $S^3_+$ represent the fact that we are using one contraction of the large $N=4$ superconformal algebra.

Similarly the superscript - on $N=4^-$ and the subscript - on $S^3_-$ represent the fact that we are using another contraction the large $N=4$ superconformal algebra.

Clearly the $N=4$ supersymmetry parts in these mappings are missing.

Finally the large $N=4$ supersymmetry of Ali-Salih's ADHM sigma model is mapped onto the large $N=4$ superconformal algebra. This in turn happens to be the superconformal symmetry of the superconformal symmetry holographically dual to superstrings moving on $AdS_ 3 \times S^3 \times S^3 \times S^1$ backgrounds.

This mappings can be represented as follows. 
\begin{center}
Complete ADHM Sigma Model

$\downarrow$ 

(Large $N=4$ SCFT)

$\downarrow $

$AdS_ 3 \times S_+^3 \times S_-^3 \times S^1$
\end{center}

Here we have differentiated between two $S^3$'s by the subscripts + and -. 

These five diagrams summarize what we have to say in this note.

\section{Discussion}\label{discussion}

In this note we have mapped the symmetries of ADHM instanton linear sigma models, $N=4$ superconformal field theories and superconformal field theories dual to  Type IIB superstrings moving on $AdS_3 \times S^3  \times K3$, $AdS_3 \times S^3  \times T^4$ and $AdS_3\times S^3\times S^3\times S^1$ backgrounds
on to each other. Corresponding supersymmetry structure should also be found in case of the two dimensional instanton sigma models 't Hooft instantons \cite{Howe:1987qv, Howe:1988cj, Strominger:1990et, Callan:1991dj, Callan:1991ky, Callan:1991at}. Indeed corresponding analysis should be simpler but we shall leave that for latter work.

The resulting structure is quite gratifying yet some features need further investigations. Foremost among these unclear features is the dynamics of infrared flow of theories with a scale, like instanton size, to conformal fixed points. The issue is compounded by the fact that all of these theories have $N=4$ supersymmetries and it is well known that these theories do not flow under renormalization (see, for a discussion in the present context, Ref.\cite{Lambert:1995dp}).

Another unsolved problem of $AdS_3$ superstrings is the fact that we still do not know the superconformal field theory that is dual to superstrings moving on $AdS_3\times S^3\times S^3\times S^1$ backgrounds completely (see Ref.\cite{Ali:2023kkf} for more specific description). We believe that the insights we get from the symmetry structures described in this note will of help in leading us to the solution of this quarter of a century old problem. 

We believe that the free field realization of super conformal field theory that is holographically dual to superstrings moving on the $AdS_ 3 \times S^3 \times S^3 \times S^1$ background geometry will help us in this problem. We are investigating this angle too.

 We also believe that the insights obtained by the corresponding analysis will help us in finding the D-brane configuration that would give above geometry in the decoupling limit. In additional we believe that the same analysis will also lead to the solution to another quarter of a century old problem of the singularity of the field theory associated with the D1/D5 system \cite{Seiberg:1999xz} because of latter's close connection with $AdS_3$ superstrings. It would be interesting to work out the highest weight representations of the large $N=4$ super conformal algebra in the light of the complete free field realization given above and find out what insights we get in addition to what we already know from \cite{deBoer:1999gea, Gukov:2004ym, Eberhardt:2017fsi} and \cite{Eberhardt:2017pty}. Of course the highest priority issue is to get to the super conformal field theory  that is holographically dual to superstrings moving on $AdS_ 3 \times S^3 \times S^3 \times S^1$ backgrounds by taking advantage of the complete free field realization  above. In a related development the off-shell superspace covariant formalism of the complementary ADHM instanton linear sigma model has been discussed in Ref.\cite{Ali:2025ntc, Ali:2025jcu}.

\textit{Acknowledgments}: Mohsin Ilahi, P.P.Abdul Salih and Shafeeq Rahman Thottoli contributed to this project during their respective Ph.D. work.

\end{document}